\documentclass[aps,prl,superscriptaddress, twocolumn, showpacs]{revtex4-1}
\usepackage{graphicx}
\usepackage{dcolumn}
\usepackage{bm}
\usepackage{amsmath}

\usepackage{color}
\usepackage{bm,graphicx,hyperref}
\hypersetup{%
  breaklinks = {true},
  citecolor = {blue},
  colorlinks = {true},
  linkcolor = {red},
}

\begin{document}

\title{Spin-Orbit Dirac Fermions in 2D Systems}

\author{A.~S.~Rodin} \affiliation{Centre for Advanced 2D
  Materials and Graphene Research Centre, National University of
  Singapore, 6 Science Drive 2, 117546, Singapore}

\author{Paul~Z.~Hanakata}
\affiliation{Department of Physics, Boston University, Boston, MA 
02215}

\author{Alexandra~Carvalho} \affiliation{Centre for Advanced 2D
  Materials and Graphene Research Centre, National University of
  Singapore, 6 Science Drive 2, 117546, Singapore}

\author{Harold~S.~Park}
\affiliation{Department of Mechanical Engineering, Boston University, Boston, MA 02215}

\author{David~K.~Campbell}
\affiliation{Department of Physics, Boston University, Boston, MA  02215}

\author{A.~H.~Castro Neto} \affiliation{Centre for Advanced 2D
  Materials and Graphene Research Centre, National University of
  Singapore, 6 Science Drive 2, 117546, Singapore}

\date{\today}
\begin{abstract}
We propose a novel model for including spin-orbit interactions in buckled two dimensional systems. Our results show that in such systems, intrinsic spin-orbit coupling leads to a formation of Dirac cones, similar to Rashba model. We explore the microscopic origins of this behaviour and confirm our results using DFT calculations.
\end{abstract}


\maketitle

\emph{Introduction.} Spin-orbit interaction (SOI) occupies a special place in physics, where even its simplest manifestation is highly non-trivial, being relativistic in nature. In solid state, SOI can be either extrinsic or intrinsic. Extrinsic system-wide SOI is generally described using Rashba formalism, where a perpendicular electric field is applied to the system, leading to a lifting of band degeneracy and a formation of Dirac cones.

Intrinsic SOI in two dimensions is limited to materials with heavy atoms, like some transition metal dichalcogenides, or graphene-like buckled Xenes. For both classes of materials, SOI results in band splitting, opening a gap~\cite{liu-PRB-84-195430-2011,liu-PRL-107-076802-2011, xu-PRL-111-136804-2013,kormanyos-PRX-4-011034-2014, Kormanyos-2DMat-2-022001-2015, molle-NatMat-16-163-2017}.

In this work, we introduce a new model for two-dimensional systems with SOI. The system considered here has a fairly simple structure and can be constructed using the existing tehcnology.~\cite{gomes-Nature-306-2012,polini-Nat-625-2013} By employing tight-binding formalism on a buckled homopolar square lattice with inversion symmetry, we show that instead of opening a gap, SOI can lead to a Rashba-like dispersion without an external field. After constructing a compact model, we use \emph{ab initio} calculations to confirm our results. 

\begin{figure}[h]
\includegraphics[width = 2.75in]{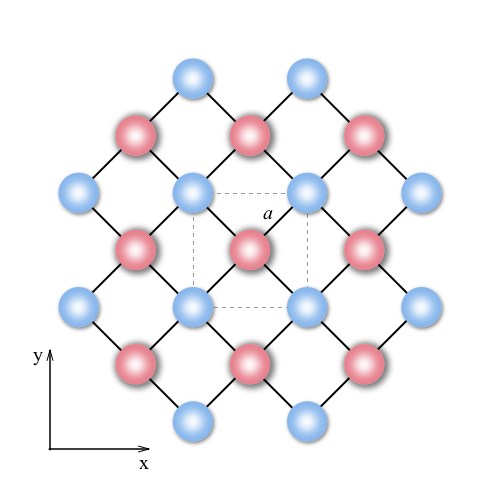}
\caption{Buckled square lattice with the two sublattices highlighted. Sublattice A is elevated above sublattice B. The dashed square marks the unit cell.}
\label{fig:Lattice}
\end{figure}

\emph{Model.} We demonstrate that SOI can
lead to Dirac dispersion in two-dimensional materials if certain
geometric requirements are satisfied. We begin our discussion by
considering a buckled square lattice, composed of a single atomic
species. Because of the buckling, we can regard the lattice as being
composed of two inequivalent shifted square sublattices A and B, see
Fig.~\ref{fig:Lattice}. We set the bond length to $a$ and the
(buckling) angle that it makes with the horizontal to $\theta$. This
yields the unit cell size of $2\alpha a\times 2\alpha a$, where
$\alpha = \cos\theta/\sqrt{2}$. From this, the dimensions of the
Brillouin zone are
$-\frac{\pi}{2\alpha a}\leq q_x,q_y\leq\frac{\pi}{2\alpha a}$.

To keep the model as simple as possible, we only consider $p$ orbitals for each atom. We will address the effects that including $s$ orbitals can have below. Given this approximation, we now construct a tight-binding Hamiltonian, allowing transport only between the nearest neighbors. Because all the atoms and, therefore, the orbitals are identical, we set the orbital on-site energy to 0. In addition, since we are limiting hopping to the nearest neighbors, there is no coupling between the atoms of the same sublattice. The hopping matrix connecting the different sublattices, recall that the coupling term between two $p$ orbitals is given by~\cite{slater-PR-94-1498-1954}
\begin{equation}
t = \left(\hat{o}_1\cdot\hat{o}_2\right) V_\pi+\left(\hat{o}_1\cdot\hat{d}\right)\left(\hat{o}_2\cdot\hat{d}\right)\left(V_\sigma-V_\pi\right)\,,
\label{eqn:Hopping_Int}
\end{equation}
where $\hat{o}_i$ is the orientation of the $i$th orbital, $\hat{d}$ is the unit vector pointing from atom 1 to atom 2, and $V_\sigma$ ($V_\pi$) is the hopping integral for the $\sigma$ ($\pi$) bonds. The vectors connecting sublattice A to its four nearest neighbors in sublattice B are $a\left(n\alpha,\,m\alpha,\,-\beta\right)$ for $m$ and $n = \pm 1$, and $\beta = \sin\theta$. Thus, the two sublattices are coupled by:
\begin{widetext}
\begin{align}
K &= 4\cos\left(k_x\right)\cos\left(k_y\right)\begin{pmatrix}
\alpha^2\Delta+V_{pp\pi}&0&0
\\
0&\alpha^2\Delta+V_{pp\pi}&0
\\
0&0&\beta^2\Delta+V_{pp\pi}
\end{pmatrix}+4\sin\left(k_x\right)\sin\left(k_y\right)\begin{pmatrix}
0&-\alpha^2\Delta&0
\\
-\alpha^2\Delta&0&0
\\
0&0&0
\end{pmatrix}
\nonumber
\\
&+4\cos\left(k_y\right)\sin\left(k_x\right)\begin{pmatrix}
0&0&-i\alpha\beta\Delta
\\
0& 0&0
\\
-i\alpha\beta\Delta&0&0
\end{pmatrix}+4\cos\left(k_x\right)\sin\left(k_y\right)\begin{pmatrix}
0&0&0
\\
0& 0&-i\alpha\beta\Delta
\\
0&-i\alpha\beta\Delta&0
\end{pmatrix}\,,
\label{eqn:K}
\end{align}
\end{widetext}
where $\Delta = V_{pp\sigma}-V_{pp\pi}$, and the momentum $k_{x,y} \in \left[-\pi/2,\pi/2\right]$.

Even though it is convenient to use $p_x$ and $p_y$ orbitals to write down the hopping matrix, since we are interested in including SOI in our model, it is helpful to go to a basis which is more natural for the angular momentum operators: $|1,1\rangle = -(|p_x\rangle+i|p_y\rangle)/\sqrt{2}$ and $|1,-1\rangle = (|p_x\rangle-i|p_y\rangle)/\sqrt{2}$. In addition, we will focus our attention on the corner of the Brillouin zone. The reason for this is that, according to Eq.~\eqref{eqn:K}, a large number of the orbitals decouple at the M point. This both simplifies our analysis and further validates our dropping of $s$ orbitals. Applying the transformation described above and expanding the trigonometric functions around $(\pi/2,\,\pi/2)$ transforms $K$ into
\begin{equation}
\frac{K}{\Delta}\rightarrow4i\alpha^2\begin{pmatrix}
0&-1&0\\1&0&0\\0&0&0
\end{pmatrix}+2\sqrt{2}\alpha\beta k\begin{pmatrix}
0&0&-e^{i\phi}\\0&0&-e^{-i\phi}\\e^{-i\phi}&e^{i\phi}&0
\end{pmatrix}\,.
\label{eqn:K_Trans}
\end{equation}
The first term couples in-plane orbitals with opposite angular momenta; the second term couples in-plane orbitals to $p_z$ at finite $k$. Note that this coupling is spin-preserving.

To add the spin effects to our model, we use the standard form describing the spin-orbit coupling:
\begin{equation}
\frac{H_\mathrm{SOI}}{\Delta} = T\left(\frac{L_+\otimes s_-+L_-\otimes s_+}{2}+L_z\otimes s_z\right)\,,
\label{eqn:SOI}
\end{equation}
where $T$ is the dimensionless ratio between the spin-orbit coupling and the hopping energy. 
The last term modifies the diagonal elements of the self-energy for
$|1,\pm1\rangle$ by adding (subtracting) $T/2$ if $L_z$ and
$s_z$ point in the same (opposite) direction. The first tem couples
$|1,1\rangle\otimes |\downarrow\rangle$ with
$|1,0\rangle\otimes|\uparrow\rangle$ and
$|1,-1\rangle\otimes |\uparrow\rangle$ with
$|1,0\rangle\otimes|\downarrow\rangle$ with the coupling strength
$T/\sqrt{2}$.

Strictly at the M point, the system decouples into four block Hamiltonians:
\begin{equation}
H_1 = \begin{pmatrix}
\frac{T}{2}&\mp4i\alpha^2&0\\\pm4i\alpha^2&-\frac{T}{2}&\frac{T}{\sqrt{2}}\\0&\frac{T}{\sqrt{2}}&0
\end{pmatrix}\,,
\end{equation}
with the top (bottom) signs acting over $|1,1\rangle\otimes|\uparrow\rangle_A$, $|1,-1\rangle\otimes|\uparrow\rangle_B$, $|1,0\rangle\otimes|\downarrow\rangle_B$ ($|1,-1\rangle\otimes|\downarrow\rangle_A$, $|1,1\rangle\otimes|\downarrow\rangle_B$, $|1,0\rangle\otimes|\uparrow\rangle_B$) and 
\begin{equation}
H_2 = \begin{pmatrix}
-\frac{T}{2}&\frac{T}{\sqrt{2}}&\mp4i\alpha^2\\\frac{T}{\sqrt{2}}&0&0\\\pm4i\alpha^2&0&\frac{T}{2}
\end{pmatrix}\,,
\end{equation}
where the top (bottom) signs operate on $|1,1\rangle\otimes|\downarrow\rangle_A$, $|1,0\rangle\otimes|\uparrow\rangle_A$, $|1,-1\rangle\otimes|\downarrow\rangle_B$ ($|1,-1\rangle\otimes|\uparrow\rangle_A$, $|1,0\rangle\otimes|\downarrow\rangle_A$, $|1,1\rangle\otimes|\uparrow\rangle_B$). 

Having constructed a simplified Hamiltonian around the M point of the square Brillouin zone, we now demonstrate that the presence of the atomic spin-orbit coupling leads to a formation of the Dirac cones. From the form of $H_1$ and $H_2$ it is clear that they result in a four-fold degenerate states at the M point. The most general form for these eigenstates is
\begin{align}
|\Psi_1^\mathrm{I}\rangle &= ia|1,1\rangle\otimes|\uparrow\rangle_A+b|1,-1\rangle\otimes|\uparrow\rangle_B+c|1,0\rangle\otimes|\downarrow\rangle_B\,,
\nonumber
\\
|\Psi_1^\mathrm{II}\rangle &= -ia|1,-1\rangle\otimes|\downarrow\rangle_A+b|1,1\rangle\otimes|\downarrow\rangle_B+c|1,0\rangle\otimes|\uparrow\rangle_B\,,
\nonumber
\\
|\Psi_2^\mathrm{I}\rangle &= ia|1,-1\rangle\otimes|\downarrow\rangle_B+b|1,1\rangle\otimes|\downarrow\rangle_A+c|1,0\rangle\otimes|\uparrow\rangle_A\,,
\nonumber
\\
|\Psi_2^\mathrm{II}\rangle &= -ia|1,1\rangle\otimes|\uparrow\rangle_B+b|1,-1\rangle\otimes|\uparrow\rangle_A+c|1,0\rangle\otimes|\downarrow\rangle_A\,,
\label{eqn:Eigenstates}
\end{align}
where $a$, $b$, and $c$ are real.

\begin{figure}
\includegraphics[width = 3in]{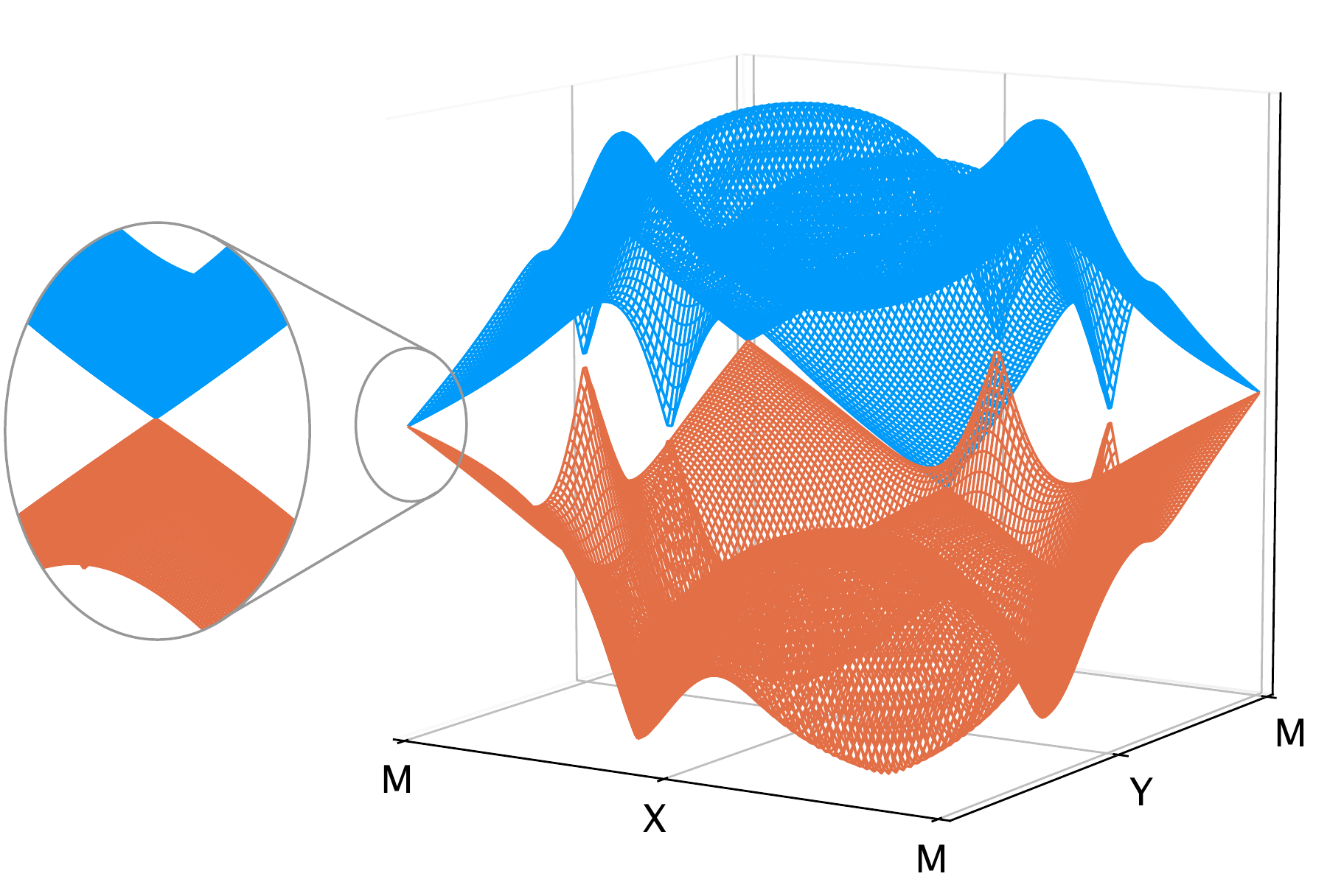}
\caption{Two out of six bands plotted over the Brillouin zone for $T = 1/2$, $\theta = \pi/12$. We set $V_\pi\rightarrow 0$ to reduce the number of parameters for illustrative purposes.}
\label{fig:Bands}
\end{figure}

Using Eq.~\eqref{eqn:K_Trans}, we get that the coupling between the degenerate pairs for $H_1$ and $H_2$ is
\begin{align}
H_2^\mathrm{eff} =\left(H_1^\mathrm{eff} \right)^*&= 2\sin2\theta ac\left[\left(\mathbf{k}\times\mathbf{\sigma}\right)\cdot\hat{z}\right]\Delta\,,
\label{eqn:Coupling}
\end{align}
which has the form of the Rashba perturbation. This results in the lifting of the degeneracy for finite values of $k$. Since the coupling between the two states is linear in momentum, the splitting forms a Dirac cone. Superimposed on the curvature of a spinless band, this linear coupling from Eq.~\eqref{eqn:Coupling} leads to a Rashba-like dispersion. On the other hand, if the spinless band is flat, the result is a well-defined Dirac cone. Figure~\ref{fig:Bands} shows two middle bands over the whole Brillouin zone for $T = 1/2$, $V_\pi = 0$, and $\theta = \pi/12$ with the linear bands at the M point.

An important feature that sets our system apart from others Dirac systems is the fact that we have only one valley. In other materials (like graphene and TMDC's), the cones appear at the inequivalent valleys at K and K' points of the Brillouin zone. Here, on the other hand, the cone appears at the unique $M$ point.

In the case of Rashba splitting, the Dirac cones describe helical
states, where in-plane spin is linked to momentum. At first glance, it
might appear that our system will behave in a similar fashion. Indeed,
the Rashba-like Hamiltonian in Eq.~\eqref{eqn:Coupling} does give rise
to helical states. However, the two Hamiltonians in
Eq.~\eqref{eqn:Coupling} have opposite helicities, resulting in a
cancellation of the spin texture. In order to avoid this cancellation
one can get rid of the equivalence of $H_1$ and $H_2$ by, for
example, replacing one of the sublattices by a different atomic
species.

Finally, we look at the pseudo-spin, as defined by Eq.~\eqref{eqn:Coupling}. The eigenstates for $H_2^\mathrm{eff}$ are $(1,\,\pm i e^{i\phi})/\sqrt{2}$. Treating the first component of the spinor as $|\uparrow\rangle$ for the pseudo-spin and the second component as $|\downarrow\rangle$, we see that the pseudo-spin forms a vortex around the origin, pointing at $90^\circ$ to the direction of the momentum. The direction of the pseudo-spin for the top and bottom cones are opposite to each other. In addition, the direction for $H_1^\mathrm{eff}$ is opposite to that of $H_1^\mathrm{eff}$, yielding no net pseudo-spin texture for the degenerate system. Interestingly, the vortex-like texture is different from what one observes in graphene, where the pseudo-spin is parallel to the momentum.

\emph{Discussion.} Having demonstrated that our simplified system
does, in fact, possess spin-split bands, let us now look at the
mechanism that leads to their formation. We start by turning our
attention to the composition of the matrix element in
Eq.~\eqref{eqn:Coupling}. We have already stated that its linear $k$
dependence results in Dirac-like bands. The second factor,
$\sin2\theta$, shows that for a flat lattice ($\theta=0$), the linear
coupling vanishes. The following factor, $ac$, indicates that the band splitting
takes place only if the degenerate states contain orbitals of opposite
spins. The only mechanism that we have which mixes spin is
SOI. Treating $T$ as a perturbation for the blocks in the first term
of $H_1$ and $H_2$ reveals that the $ac$ term is proportional to $T$
for small $T$. Combined with the $\Delta$ factor, this means that in
its leading-order behavior, the matrix element in
Eq.~\eqref{eqn:Coupling} is proportional to $T\Delta$. Unlike Rashba
effect, where the coupling element is always linear in the spin-orbit
term, here, increasing $T$ yields a non-linear dependence of
Eq.~\eqref{eqn:Coupling} on the ratio between the spin-orbit and
hopping terms.

\begin{figure}
\includegraphics[width = 3in]{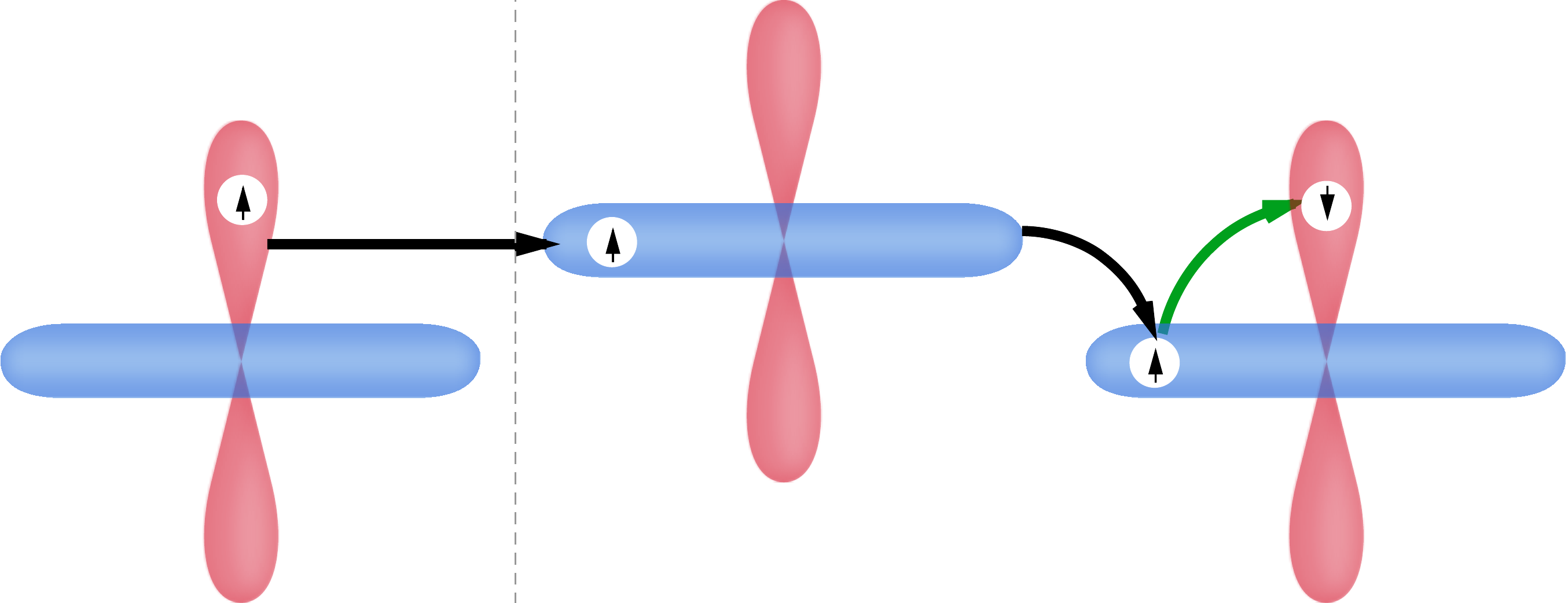}
\caption{Hopping path, leading to a spin-flip for an out-of-plane orbital. The first and third atom belong to the same sublattice, the middle one is a part of the other sublattice. The transition between the out-of-plane and in-plane orbitals is only allowed if the lattice is buckled. The transitions on the right hand side of the dashed line take place within one eigenstate (e.g., $|\Psi_1^\mathbf{II}\rangle$). The hop across the dashed line is the transition between the two degenerate eigenstates of the Hamiltonian.}
\label{fig:SOC_Hopping}
\end{figure}

As we said in the introduction, the goal of this paper is to gain
understanding of the effects of spin-orbit coupling on microscopic
level. To do so, we focus on the composition of the states in
Eq.~\eqref{eqn:Eigenstates}. We can see that both of them include
$|1,0\rangle_B$, but with the opposite spin. The fact that these
states are coupled means that there is a finite amplitude for
spin-flipping processes for
$|1,0\rangle$. Figure~\ref{fig:SOC_Hopping} shows sets of steps for
changing the spin for one of the sublattices. There are several
important features in the diagram, so let us address them
carefully. First, the illustration makes it clear why the buckling in
the system is crucial. The path involves a transition between in-plane
and out-of-plane orbitals. For a flat lattice, $p_z$ orbitals decouple
from $p_x$ and $p_y$, making these hops impossible. This is in
agreement with the $\sin2\theta$ term in
Eq.~\eqref{eqn:Coupling}. Next, the path involves only one
SOI-mediated transition. In other words, the spin-flipping process is
first-order in $T$. Of course, as more complex paths are added, higher
powers of $T$ will be included. This is equivalent to going to higher
order perturbation terms in the expansion of $ac$ in
Eq.~\eqref{eqn:Coupling}. The atoms on the right of the vertical
dashed line all belong to the same state in
Eq.~\eqref{eqn:Eigenstates}. The first hop is the transition between
the states and the amplitude of this transition is proportional to the
product of the relevant orbital amplitudes $ac$, as expected.

From our cartoon illustration in Fig.~\ref{fig:SOC_Hopping}, it might
appear that we have done unnecessary work by considering a square
lattice because one only needs a zigzag 1D chain, similar to
polyacetylene. It is possible to show that 1D chains are not
sufficient by performing a full tight-binding analysis on such a
chain. However, an easier way to see that 1D chain is insufficient
involves rotating it $90^\circ$ around the longitudinal axis so that
the zigzags are in the $xy$ plane. In this orientation, $p_z$ orbitals
are decoupled and, as we determined above, one needs coupling between
in-plane and out-of-plane orbitals to observe the SOI-induced band
splitting.

For the sake of
completeness, let us analyze the effects that including $s$
orbitals would have. If one were to construct a spinful Hamiltonian
around $\Gamma$ point, similarly to how we did it around M point, one
could see that the end result would not include the degenerate states,
coupled at finite $k$. The two equivalent Hamiltonian blocks, while
degenerate, would still be uncoupled. This means that our model would
not have Dirac cones at the $\Gamma$ point. Now, we include $s$
orbitals. Because of their symmetry, $s$ orbitals couple to $p_z$
orbitals of the other sublattice if the lattice is buckled. This would
enable the coupling between the Hamiltonian blocks and lift the
degeneracy. It is important to keep in mind that because of the energy
difference between $s$ and $p$ orbitals, the eigenstates will be
predominantly one or the other. Since the magnitude of the band
splitting depends on the presence of both orbital types, it will
generally be weaker than what's expected at the M point.

\emph{First Principles Calculations.}  We
performed density functional theory (DFT) calculations implemented in
{\sc Quantum ESPRESSO} package~\cite{QE-2009} to simulate heavy
elements Pb and Sn in a square lattice geometry. We employed Projector
Augmeneted-Wave (PAW) pseudopotential with Perdew-Burke-Ernzerhof (PBE)
for the exchange and correlation functional within the generalized
gradient approximation (GGA)~\cite{pbe-PRL-77-3865-1996}. The
Kohn-Sham orbitals were expanded in a plane-wave basis with a cutoff
energy of 70 Ry, and for the charge density a cutoff of 280 Ry was
used.  A $k$-point grid sampling grid was generated using the
Monkhorst-Pack scheme with 16$\times$16$\times$1
points~\cite{monkhorst-PRB-13-5188-1976}, and a finer regular grid of
40$\times$40$\times$1 was used for spin texture calculations. For
electronic band structure calculations, the spin orbit interaction was
included using noncollinear calculations with fully relativistic
pseudopotentials.

\begin{figure}[h]
\includegraphics[width = 3in]{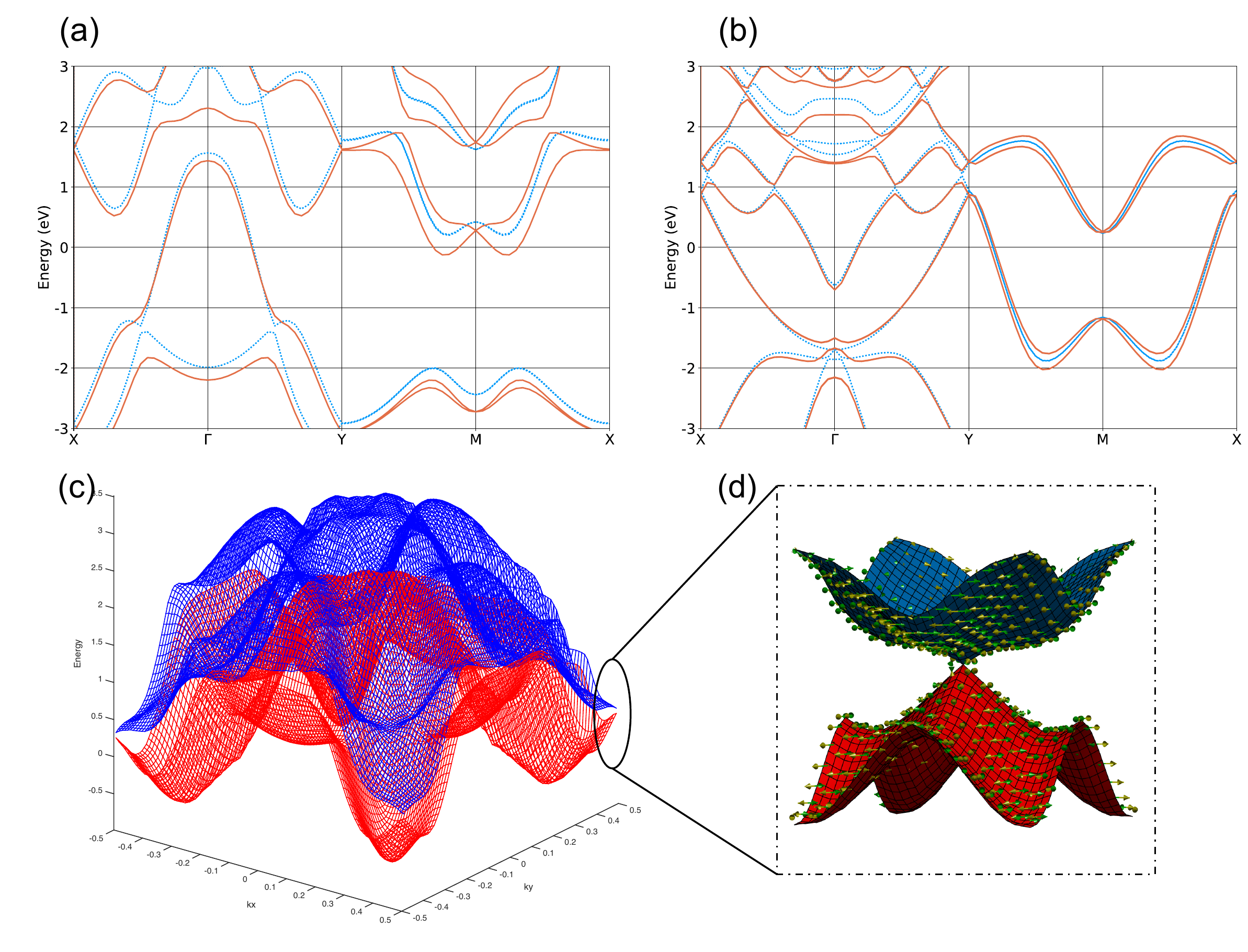}
\caption{Band structures of Pb (a) and Sn (b). Red lines indicate when
  spin-orbit interaction is included. (c) Energy surface over
  Brillouin zone of Pb monolayer and (d) its Dirac cone at M with spin
  texture.}
\label{fig:DFT_Bands}
\end{figure}

The optimized lattice constant $2\alpha a$ and buckling angle $\theta$
for Pb (Sn) are 3.44 \AA (3.83 \AA) and 44.3$^{\circ}$
(27.4$^{\circ}$), respectively. We found Dirac cone or Rashba-like
dispersion near $M$-point which are consistent with our TB
prediction, shown in Fig.~\ref{fig:DFT_Bands}. As expected from
Eq.~\eqref{eqn:Coupling}, the lighter element Sn has a smaller Dirac
slope due to smaller value of $T$ and buckling angle $\theta$.

The top and bottom Dirac cone are colored blue and red,
respectively. Each band is doubly degenerate with opposite spin
texture resulting a zero net spin texture. Recently, it has been
proposed that centrosymmetric materials might have spin
polarization~\cite{zhang-NatPhys-10-387-2014}. It was found that the
top and the bottom sector of LaOBiS$_2$ have opposite spin
texture~\cite{zhang-NatPhys-10-387-2014}. Our works support such
claims as Pb and Sn monolayers are centrosymmetric
materials.

\emph{Conclusions} We have demonstrated that intrinsic spin-orbit
coupling in buckled 2D materials can lead to the appearance of the
Dirac cones. Even though the effective Hamiltonian and the band
structure is reminiscent of the well-known Rashba effect, the
underlying physics is fundamentally different. In our study, we
focused on a square lattice, but the results hold for any 2D material
where in-plane orbitals couple to out-of-plane ones.


%

\end{document}